\documentstyle[11pt]{article}

\def\dalemb#1#2{{\vbox{\hrule height .#2pt
        \hbox{\vrule width.#2pt height#1pt \kern#1pt
                \vrule width.#2pt}
        \hrule height.#2pt}}}

\def\td{\tilde}

\let\a=\alpha

 \def\bd{\begin{document}} \def\ed{\end{document}}
\def\ds{\documentstyle} \let\fr=\frac \let\bl=\bigl \let\br=\bigr
\let\Br=\Bigr \let\Bl=\Bigl
\let\bm=\bibitem
\let\na=\nabla
\let\pa=\partial \let\ov=\overline
\newcommand{\be}{\begin{equation}}
\newcommand{\ee}{\end{equation}}
\def\ba{\begin{array}}
\def\ea{\end{array}}
\def\ft#1#2{{\textstyle{{\scriptstyle #1}\over {\scriptstyle #2}}}}
\def\fft#1#2{{#1 \over #2}}
\def\del{\partial}
\def\sst#1{{\scriptscriptstyle #1}}
\def\oneone{\rlap 1\mkern4mu{\rm l}}
\newcommand{\ho}[1]{$\, ^{#1}$}
\newcommand{\hoch}[1]{$\, ^{#1}$}
\newcommand{\bea}{\begin{eqnarray}}
\newcommand{\eea}{\end{eqnarray}}
\newcommand{\ra}{\rightarrow}
\newcommand{\lra}{\longrightarrow}
\newcommand{\Lra}{\Leftrightarrow}
\newcommand{\ap}{\alpha^\prime}
\newcommand{\bp}{\tilde \beta^\prime}
\newcommand{\tr}{{\rm tr} }
\newcommand{\Tr}{{\rm Tr} }
\newcommand{\NP}{Nucl. Phys. }
\newcommand{\tamphys}{\it Center for Theoretical Physics,
Texas A\&M University, College Station, Texas 77843}

\newcommand{\auth}{H. L\"u and C.N. Pope}

\begin{document}

\hfill{CTP-TAMU-52/95}

\hfill{hep-th/9512153}

\vspace{20pt}

\begin{center}
{ \large {\bf Multi-Scalar $p$-brane Solitons }}

\vspace{30pt}
\auth

\vspace{15pt}
{\tamphys}
\vspace{40pt}

\underline{ABSTRACT}
\end{center}
\vspace{40pt}

  In a previous paper \cite{lp}, supersymmetric $p$-brane solutions
involving one dilatonic scalar field in maximal supergravity theories were
classified.  Although these solutions involve a number of participating
field strengths, they are all equal and thus they carry equal electric or
magnetic charges.  In this paper, we generalise all these solutions to
multi-scalar solutions in which the charges become independent free
parameters.  The mass per unit $p$-volume is equal to the sum of these
Page charges.  We find that for generic values of the Page charges, they
preserve the same fraction of the supersymmetry as in their single-scalar
limits. However, for special values of the Page charges, the supersymmetry
can be enhanced.

{\vfill\leftline{}\vfill
\vskip	10pt
\footnoterule
{\footnotesize	Research supported in part by DOE
Grant DE-FG05-91-ER40633 \vskip	-12pt}}

\pagebreak
\setcounter{page}{1}

      Isotropic $p$-brane solitons from supergravity theories have been
extensively studied in recent years.  Most of the solutions that have been
found can be described in terms of a single dilatonic scalar field and a
single antisymmetric tensor field strength [1-10].  Two possible kinds of
solution arise, one carrying an ``electric'' charge for the field strength,
and describing a fundamental or elementary $p$-brane, and the other carrying
a ``magnetic'' charge, corresponding to a solitonic $p$-brane solution.
(When the dimension is twice the degree of the field strength, dyonic
solutions that carry both electric and magnetic charges can also arise.) In
dimensions $D$ lower than 10, the scalar field might be a linear combination
of the dilatonic scalar fields in the maximal supergravity theory in that
dimension.  At the same time, more than one of the original antisymmetric
tensor fields might be non-zero in the solution, although all of them will
be proportional to one another.  Thus there is only one overall charge
parameter characterising any given solution, with the electric or magnetic
charges of the individual field strengths occurring in a fixed ratio.

     Recently, some further solutions have been found in which the charges
of participating field strengths are independent of one another [11-14].
This is achieved by relaxing the condition that only one linear combination
of the dilatonic scalar fields is non-vanishing, and so these solutions may
be characterised as multi-scalar solutions.  Each of the previous single
scalar solutions with more than one participating field strength therefore
has such a multi-scalar generalisation. In this paper, we shall give a
systematic construction of multi-scalar supersymmetric solutions in
maximal supergravity theories in $4\le D \le 9$.  The supergravity theories
that we shall consider are those that are obtained from type IIA
supergravity in $D=10$, or, equivalently, from $D=11$ supergravity, by
Kaluza-Klein dimensional reduction. For convenience, we shall consider the
supergravity theories in the versions where all the field strengths have
degrees $n$ that are less than or equal to 4.  Some related results for
0-branes in $D=4$ have previously been obtained in [11-14], mostly in the
context of the 4-dimensional heterotic string.

     Let us consider a $p$-brane solution in which a number $N$ of $n$-index
($n=1,2,3,4$) field strengths $F^\a$ ($\a = 1,2, \ldots, N$) are involved. The
relevant part of the bosonic Lagrangian for the supergravity theory is given by
\be
e^{-1}{\cal L} = R -\ft12 (\del \vec \phi)^2 -\fft{1}{2n!} \sum_{\a}^N
e^{\vec a_\a \cdot \vec \phi} (F^\a)^2\ ,\label{lag1}
\ee
where $\vec \phi = (\phi_1, \phi_2, \ldots, \phi_{11-D})$ are the $(11-D)$
dilatonic scalar fields.  The dilatonic vectors $\vec a_\a$ follow by
dimensional reduction from $D=11$ supergravity, and can be found, for
example, in \cite{lp}.  In general, there are further contributions in the
bosonic Lagrangian coming from the dimensional reduction of $F\wedge F\wedge
A$ in $D=11$ and from the Chern-Simons modifications that the field
strengths acquire in the dimensional reduction process.  We shall be
concerned only with solutions where these terms do not contribute.  This
imposes certain constraints on the field configurations, which were
discussed in \cite{lp}.

    The metric ansatz is given by
\be
ds^2 = e^{2A} dx^\mu dx^\nu \eta_{\mu\nu} + e^{2B} dy^mdy^m\ ,
\label{metricform}
\ee
where $x^\mu (\mu = 0,\ldots, d-1)$ are the coordinates of the
$(d-1)$-brane volume, and $y^m$ are the coordinates of the
$(D-d)$-dimensional transverse space.  The function $A$ and $B$, as well as
all the dilatonic scalars $\vec \phi$, depend only on $r=\sqrt{y^my^m}$.
Thus the ansatz preserves an $SO(1,d-1)\times SO(D-d)$ subgroup of the
original $SO(1,D-1)$ Lorentz group.

     For each $n$-index field strength $F^\a$, there are two different
ans\"atze that also preserve the same subgroup, namely \cite{dabl,pew2}
\be
F^\a_{m\mu_1\cdots\mu_{n-1}} = \epsilon_{\mu_1\cdots\mu_{n-1}} (e^{C_\a})'\,
\fft{y^m}{r}\qquad {\rm or}\qquad
F^\a_{m_1\cdots m_n }= \lambda_\a \,\epsilon_{m_1\cdots m_n p}\,
\fft{y^p}{r^{n+1}}\ ,\label{fansatz}
\ee
where a prime denotes a derivative with respect to $r$.  The first case
gives rise to an elementary $(d-1)$-brane with $d=n-1$ for $n=2,3$ or 4; the
second gives rise to a solitonic $(d-1)$-brane with $d=D-n-1$ for
$n=1,2,3$ or 4.  In this paper, we consider solutions with each $F^\a$
having either elementary or solitonic contributions, but not both.  In
$D=2n$, some $F^\a$ might be the duals of the original field strengths of
the same degree.  Thus in terms of the original field strengths, such
solutions have both elementary and solitonic contributions, giving rise to
dyonic solutions of the first type \cite{lp}.  As we shall see later, these
dyonic solutions are possible in $D=4$, but not in $D=8$ or $D=6$.

     Let us first summarise the known results for single-scalar $p$-brane
solutions.  In \cite{lp}, the equations of motion following from (\ref{lag1})
were solved by first truncating to a single field strength and a single
scalar field, with Lagrangian:
\be
e^{-1}{\cal L} = R -\ft12 (\del \phi)^2 -\fft{1}{2n!} e^{a\phi} F^2\ ,
\label{lag2}
\ee
where $\phi$ is a linear combination of the original dilatonic scalars $\vec
\phi$, and $F$ is the canonically-normalised field strength formed from the
participating field strengths $F^\a$.  In order to be able to set the
orthogonal combinations of dilatonic scalars to zero, in a manner consistent
with their equations of motion, the field strengths $F^\a$ must be all
proportional to $F$ \cite{lp}: In the generic case, where the matrix
$M_{\a\beta} \equiv \vec a_\a \cdot \vec a_\beta$ is non-singular, we must
have
\bea
a^2 &=& \Big (\sum_{\a,\beta} (M^{-1})_{\a\beta}\Big)^{-1}\ ,\qquad
\phi = a \sum_{\a,\beta} (M^{-1})_{\a\beta}\, \vec a_\a \cdot \vec\phi
\ ,\nonumber\\
(F^\a)^2 &=& a^2 \sum_\beta (M^{-1})_{\a\beta}\, F^2\ .\label{single}
\eea
When the matrix $M_{\a\beta}$ is singular, one finds that the only new
solution that need be considered is for the case $\sum_\a \vec a_\a=0$,
which gives rise to $a=0$ and $(F^\a)^2=F^2/N$ for all $\a$ \cite{lp}.
It is convenient to parameterise the dilaton prefactor $a$ by \cite{lpss}
\be
a^2 = \Delta - \fft{2d\td d}{D-2}\ ,\label{avalue}
\ee
where $\tilde d = D-d-2$ and $d \td d = (n-1)(D-n-1)$.  It is then
straightforward to solve the equations of motion following from the
Lagrangian (\ref{lag2}), using the metric ansatz (\ref{metricform}) and
field strength ans\"atze (\ref{fansatz}).   The metrics of the solutions for
given degree of the field strength in dimension $D$ are determined by the
value of $\Delta$ \cite{lpss}, and are given by
\be
ds^2 = \Big(1 + \fft{k}{r^{\td d}}\Big)^{-\fft{4\td d}{\Delta (D-2)}}\,
dx^\mu dx^\nu \eta_{\mu\nu} +
\Big(1 + \fft{k}{r^{\td d}}\Big)^{\fft{4 d}{\Delta (D-2)}} \, dy^m dy^m
\ ,\label{metricsol1}
\ee
where $k =\sqrt{\Delta} \lambda/(2\td d)$.  The mass per unit $p$-brane
volume is given by $m=\ft12 (B'-A')e^{-B} r^{\td d +1}$ in the limit $r
\longrightarrow \infty$ \cite{lp}.  The Page charge $P$ \cite{page} for the
canonically-normalised field strength $F$ is given by $\ft1{4 \omega_{\sst D
-n}} \int_{S^{(\sst D -n)}} *F$ for the elementary case and by
$\ft1{4\omega_n} \int_{S^n} F$ for the solitonic case.  Thus we have
\be
m= \fft{\lambda}{2\sqrt{\Delta}}\ ,\qquad P = \ft14 \lambda\ .\label{mc1}
\ee
Note that the Page charge $P_\a$ of each individual field strength $F^\a$
is a certain fixed multiple of $P$ in a given solution, as determined by
(\ref{single}).

    Now we turn to the consideration of multi-scalar solutions. Substituting
the ans\"atze (\ref{fansatz}) and (\ref{metricform}) directly into the
equations of motion that follow from the Lagrangian (\ref{lag1}), we find
that $\vec \phi$, $A$ and $B$ satisfy
\bea
&&\vec \phi'' +\fft{\td d +1}{r}\, \vec \phi' +(dA' + \td d B')\vec \phi'
= -\ft12\epsilon \sum_\a \vec a_\a\, S^2_\a\ ,\label{eom11}\\
&&A'' + \fft{\td d +1}{r}\, A' + (dA' + \td d B') A' =
\fft{\td d}{2(D-2)} \sum_\a S_\a^2\ ,\label{eom12}\\
&&B'' +\fft{\td d +1}{r}\, B' + (dA' + \td d B') (B' + \fft1{r})
=-\fft{d}{2(D-2)} \sum_a S_\a^2\ ,\label{eom13}\\
&&d(D-2) A'^2 + \td d (d A'' + \td d B'') - (d A' + \td d B')^2
-\fft{\td d}{r} (d A' + \td d B') + \ft12 \td d \, \vec\phi'^2
= \ft12 \td d \sum_\a S_\a^2\ ,\label{eom14}
\eea
where $\epsilon =1$ and $-1$ for the elementary and solitonic ans\"atze
respectively, and the functions $S_\a$ are given by
\be
S_\a = \lambda_\a\, e^{-\ft12\epsilon \vec a_\a\cdot \vec \phi
-\td d B}\, r^{-\td d -1}\ .\label{salpha}
\ee
In the elementary case, $\lambda_\a$ arises as the integration constant for
the function $C$, given by
\be
(e^{C_\a})' = \lambda_\a \, e^{\vec a_\a \cdot \vec \phi + d A - \td B}\,
r^{-\td d -1}\ .
\ee

      From (\ref{eom12}) and (\ref{eom13}), we see that a natural solution
for $B$ is to take
\be
d A + \td d B=0\ .
\ee
We may also consistently set to zero the
$(11-D-N)$ components of $\vec \phi$ that are
orthogonal to the space spanned by the $N$ dilaton vectors $\vec a_\a$.  The
remaining equations are
\bea
&&\varphi_\a'' + \fft{\td d +1}{r}\, \varphi_\a' =
-\ft12 \epsilon \sum_\beta M_{\a\beta}\, S^2_\beta\ ,\label{eom21}\\
&&A'' + \fft{\td d +1}{r} \, A' = \fft{\td d}{2(D-2)} \sum_\a S_\a^2
\ ,\label{eom22}\\
&&d(D-2) A'^2 +\ft12 \td d \sum_{\a,\beta} (M^{-1})_{\a\beta}\, \varphi_\a'
\varphi_\beta' =\ft12 \td d \sum_\a S_\a^2\ ,\label{eom23}
\eea
where we have defined $\varphi_\a = \vec a_\a \cdot \vec \phi$. (Here we are
assuming that $M_{\a\beta}$ is non-singular, and we shall comment on the
case when it is singular later.) Note that the number of non-vanishing
scalar fields $\varphi_\a$ is precisely the same as the number $N$ of
participating field strengths. By acting on (\ref{eom21}) with
$(M^{-1})_{\a\beta}$, and comparing with (\ref{eom22}), we see that it is
natural to solve for $A$ by taking
\be
A =- \fft{\epsilon \td d}{D-2}
\sum_{\a,\beta} (M^{-1})_{\a\beta}\, \varphi_\a\ .
\ee
The equations of motion now reduce to
\bea
\sum_{\beta} (M^{-1})_{\a\beta}\Big(\varphi_\beta'' +
\fft{\td d +1}{r}\, \varphi_\beta'\Big)
&=& -\ft12 \epsilon \lambda^2_\a\,
e^{-\epsilon\varphi_\a + 2d A}\, r^{-2(\td d +1)}\ ,
\label{eom31}\\
d(D-2) A'^2 + \ft12 \td d \sum_{\a,\beta} (M^{-1})_{\a\beta} \,
\varphi_\a'\varphi_\beta' &=&\ft12 \td d \sum_\a \lambda_\a^2 \,e^{-\epsilon
\varphi_\a + 2d A}\, r^{-2(\td d+1)}\ .\label{eom32}
\eea
As in the case of the solutions that involve only one dilatonic scalar
field, the solutions here are determined completely by the structure of the
dot products $M_{\a\beta}$ of dilaton vectors $\vec a_\a$ of the
corresponding field strengths $F^\a$.  Solutions exist only for $N\le
(11-D)$.  In general, the solutions of (\ref{eom31}) and (\ref{eom32}) are
still very complicated.   However, we can find simple solutions if we make
the ansatz that the quantity $(-\epsilon \varphi_\a + 2d A)$ appearing in
the exponential in $S_\a^2$ is proportional to the quantity $\sum_{\beta}
(M^{-1})_{\a\beta}\, \varphi_\beta$ appearing on the left-hand side of
(\ref{eom31}).  For this to be true, it implies that $M_{\a\beta}$ must take
the form
\be
M_{\a\beta} = 4 \delta_{\a\beta} - \fft{2d\td d}{D-2}\ .\label{mmatrix}
\ee
Note that the coefficient of $\delta_{\a\beta}$ can {\it a priori} be any
constant, but it is fixed to be 4 in maximal supergravity theories, since
all the dilaton vectors in such theories have magnitude $a$ given by
(\ref{avalue}) with $\Delta=4$ \cite{lpss}.  We can now solve
(\ref{eom31}) and (\ref{eom32}) completely by making the further ansatz that
$S_\a \propto (-\epsilon \varphi_\a' + 2d A')$. The solution is given by
\bea
&&e^{\ft12 \epsilon \varphi_\a - d A} = 1 + \fft{\lambda_\a}{\td d} r^{-\td
d}\ ,\nonumber\\
&&ds^2=\prod_{\a=1}^{N} \Big(1+ \fft{\lambda_\a}{\td d}
r^{-\td d}\Big)^{-\ft{\tilde d}{(D-2)}}\, dx^\mu dx^\nu \eta_{\mu\nu} +
\prod_{\a=1}^N \Big(1+ \fft{\lambda_\a}{\td d}
r^{-\td d}\Big)^{\ft{d}{(D-2)}}\, dy^m dy^m
\ .\label{gensol}
\eea
We may now calculate the mass per unit $p$-brane volume and the Page
charges for the solution, finding
\be
m = \ft14\sum_{\a=1}^N \lambda_\a\ ,\qquad P_\a = \ft14 \lambda_\a
\ .\label{mc2}
\ee
Note that in our derivation of the solutions, we assumed that the matrix
$M_{\a\beta}$ is non-singular, and indeed the matrix given by
(\ref{mmatrix}) is non-singular in general.  However, it can be singular in
two relevant cases, namely $D=5$, $N=3$ and $D=4$, $N=4$ for the 2-form
field strengths. In these cases, the analysis requires modification;
however, it turns out that (\ref{gensol}) continues to solve the equations
of motion.

    Having obtained the generic multi-scalar solutions for matrices
$M_{\a\beta}$ satisfying (\ref{mmatrix}), it is a simple matter to search
among the dilaton vectors $\vec a_\a$ in all maximal supergravity theories
for sets that have this required form of inner product.  The selection of
the field strengths must also satisfy the constraints imposed both by the
terms coming from the dimension reduction of the $F\wedge F\wedge A$ term in
$D=11$, and by the Chern-Simons modifications to the field strengths. This
problem has in fact been solved in \cite{lp}, where single-scalar solutions
for maximal supergravities were extensively studied.  In particular, the
supersymmetric solutions were classified in \cite{lp}, and all these
solutions have $M_{\a\beta}$ satisfying (\ref{mmatrix}).  Therefore, the
multi-scalar solutions we have obtained in this paper are generalisations of
the supersymmetric single-scalar solutions involving $N\ge2$ participating
field strengths, in which the Page charges of the individual field strengths
are allowed to become independent free parameters.  It is easy to verify
that these multi-scalar solutions (\ref{gensol},\ref{mc2}) reduce to the
single-scalar solutions (\ref{metricsol1},\ref{mc1}) when the Page charges
are given by
\be
\lambda_\a = \fft{\lambda}{\sqrt N}\ ,\qquad {\rm for}\,\,{\rm all}
\,\,\a\ .
\ee

     Having established that the multi-scalar solutions that we have
obtained are generalisations of the supersymmetric single-scalar solutions,
it is of interest to examine their supersymmetry properties.  We shall
discuss this separately for field strengths of each degree $n=1,2,3,4$.
Note that in $D<2n$ dimensions, the $n$-form field strength can be dualised
to a lower degree $(D-n)$-form field strength.  We shall always do this.
Since we have shown that there is a one-to-one correspondence between the
supersymmetric single-scalar solutions and their multi-scalar
generalisations, we can classify a multi-scalar solution by the
single-scalar case that it degenerates to when the individual Page charges
are set equal.  Thus we may refer to \cite{lp} for a detailed classification
of the supersymmetric single-scalar solutions.  In this paper, we shall
study the supersymmetry properties when the Page charges become independent.

    There is only one 4-form field strength in any maximal supergravity
theory, and thus there are no multi-scalar generalisations in this case. In
fact, single-scalar solutions with only one participating field strength, of
any degree, all preserve $\ft12$ of the supersymmetry, and they admit no
multi-scalar generalisations. For 3-form field strengths, as shown in
\cite{lp}, the corresponding inner-product matrices $M_{\a\beta}$ are given
by
\be
M_{\a\beta} = 2 \delta_{\a\beta} - \fft{2(D-6)}{D-2}\ .
\ee
Thus there are no multi-scalar solutions of the type we are considering for
3-form field strengths either, since this is not of the form given by
(\ref{mmatrix}).  In fact, single-scalar solutions involving more than one
3-form field strength are all non-supersymmetric \cite{lp}.

    Let us now consider multi-scalar solutions with 2-form field
strengths, which give rise to elementary 0-branes or solitonic
$(D-4)$-branes in $D$ dimensions.  In this case, the number of participating
field strengths in supersymmetric single-scalar solutions can be $N=1,2,3$
or 4, and they occur in dimensions $D\le 10,9,5$ and 4 respectively.  In
other words, for the 2-form field strengths, only these numbers $N$ of
dilaton vectors $\vec a_\a$ can give rise to matrices $M_{\a\beta}$ of the
form given in (\ref{mmatrix}).  Thus there are multi-scalar solutions with
$N=2,3$ or 4 non-vanishing scalar fields and independent Page charges.   The
supersymmetry of these solutions can be studied using the method described
in \cite{lp,dlps}, namely by constructing the Bogomol'nyi matrix ${\cal M}$
from the Nester form for the supergravity theory.  This matrix arises from
the commutator of the conserved supercharges, and its zero eigenvalues
correspond to unbroken components of $D=11$ supersymmetry. From the general
results in \cite{lp}, the relevant terms in the Bogomol'nyi matrix for
2-form field strengths are given by
\be
{\cal M} = m\oneone + \ft12 u_{ij} \Gamma_{0ij} + p_i \Gamma_{0i} +
\ft12 v_{ij} \Gamma_{\hat1\hat2\hat3 ij} + q_i \Gamma_{\hat1\hat2\hat3i}
\ ,\label{bog2form}
\ee
where $u_{ij}$ and $p_i$ are the electric Page charges for the 2-forms
$F_{\sst{MN}ij}$ and ${\cal F}_{\sst{MN}}^{(i)}$, coming from the
dimensional reduction of the eleven-dimensional 4-form and vielbein
respectively.  Similarly, $v_{ij}$ and $q_i$ are the corresponding magnetic
Page charges.  In (\ref{bog2form}), the 0 index denotes the time coordinate
of the (elementary) 0-brane, the hatted indices run over the transverse
space of the $y^m$ coordinates of the (solitonic) $(D-4)$-brane, and the
$i,j,k$ indices run over the directions that are compactified in the
Kaluza-Klein reduction from eleven dimensions to $D$-dimensional maximal
supergravity.

     There is in general more than one way to select a set of 2-form field
strengths whose dilaton vectors satisfy (\ref{mmatrix}); however, the
supersymmetry properties for the different choices are identical, and hence
we shall present only one representative in each case.   We find that the
(elementary) Page charges, and the eigenvalues of the Bogomol'nyi matrix for
the multi-scalar solutions, calculated from (\ref{bog2form}),  are given by
\bea
N=2:&& \{p_1, u_{12}\} = \ft14 \{\lambda_1, \lambda_2 \}\ ,
\qquad {\rm for}\,\, D\le 9\ ,\nonumber\\
\mu&=&\ft12 \{ 0, \lambda_1, \lambda_2, \lambda_1 + \lambda_2 \}
\ ,\nonumber\\
N=3:&& \{u_{12}, u_{34}, u_{56} \} = \ft14 \{ \lambda_1, \lambda_2,
\lambda_3 \}\ ,\qquad {\rm for}\,\, D\le 5\ ,\nonumber\\
\mu &=& \ft12 \{ 0, \lambda_1, \lambda_2, \lambda_3, \lambda_1 + \lambda_2,
\lambda_1 + \lambda_3, \lambda_2 + \lambda_3, \lambda_1 + \lambda_2 +
\lambda_3 \}\ ,\label{eigenvalues2f}\\
N=4:&& \{u_{12}, u_{34}, u_{56}, p_7^*\} = \ft14 \{\lambda_1,\lambda_2,
\lambda_3, \lambda_4\}\ ,\qquad {\rm for}\,\, D= 4\ ,\nonumber\\
\mu&=&\ft12 \{ 0, \lambda_1+\lambda_4, \lambda_2 + \lambda_4, \lambda_3 +
\lambda_4, \lambda_1 + \lambda_2,
\lambda_1 + \lambda_3, \lambda_2 + \lambda_3, \lambda_1 + \lambda_2 +
\lambda_3 +\lambda_4\}\ .\nonumber
\eea
Here a $^*$ on a Page charge indicates that the associated field strength is
dualised. Thus $p_7^*$ is the electric charge of the dualised field strength
$*{\cal F}_{\sst{MN}}^{(7)}$, and so it is the magnetic charge in terms of
the original undualised field strength ${\cal F}_{\sst{MN}}^{(7)}$.  In
other words, it corresponds to a contribution $p_7^*
\Gamma_{\hat1\hat2\hat37}$ in the Bogomol'nyi matrix.  We shall discuss this
type of dyonic solution later.

       As we discussed earlier, the Bogomol'nyi matrix is obtained from the
commutator of the conserved {\it Hermitian} supercharges, and so its
eigenvalues should always be non-negative at the quantum level.   However,
as classical solutions, where the $\lambda_\a$'s are just free parameters or
integration constants, it is clear from (\ref{eigenvalues2f}) that the
eigenvalues will be negative for certain choices of these parameters. In
these cases, the quantum positivity argument evidently breaks down, and
hence such configurations would be disallowed at the quantum level. Thus we
should presumably restrict the choices of parameters so that all the
eigenvalues are non-negative.  Note that the last entry in the list of
eigenvalues is twice the mass of the solution for all the three cases, and
the restriction will rule out $p$-brane solutions where the charges are
chosen to make the mass equal to zero.   In fact, even if this mass is
chosen to be positive, there still can be negative eigenvalues under certain
circumstances.

    Having obtained the Bogomol'nyi matrices of the multi-scalar solutions
for 2-forms, it is now straightforward to analyse their supersymmetry,
since the zero eigenvalues correspond to unbroken components of
$D=11$ supersymmetry.  In each of the three cases in (\ref{eigenvalues2f}),
the degeneracies of each eigenvalue are equal, with the total number of
eigenvalues being 32.   Thus for generic values of the parameters
$\lambda_\a$, these 2-scalar, 3-scalar and 4-scalar solutions preserve
$\ft14$, $\ft18$ and $\ft18$ of the supersymmetry respectively. It is easy
to see that when the $\lambda_\a$ are chosen to be equal, we recover the
single-scalar solutions with $\Delta = \ft4{N}$, preserving the same
fractions of the supersymmetry as in the generic cases.

     There is a supersymmetry enhancement for certain choices of the
parameters.  First of all, we note that this occurs when any of the
parameters $\lambda_\a$ is zero, but this corresponds merely to reducing
the number of participating field strengths and scalars.  Thus we shall
assume all the parameters $\lambda_\a$ are non-zero.   In all the three
cases in (\ref{eigenvalues2f}),  there is supersymmetry enhancement when the
mass $m=\ft14 \sum_\a \lambda_\a$ is zero; however, this implies that some
of the eigenvalues are negative.  In fact there can be no supersymmetry
enhancement, while still requiring that all the eigenvalues be non-negative,
for the cases $N=2$ and 3.   The situation is different for $N=4$, and we
can have three inequivalent enhancements, given by
\bea
\lambda_1=-\lambda\ ,&&\lambda_2 =\lambda\ ,\qquad
\mu = \ft12\{ 0_8, (\lambda_3\pm\lambda)_4, (\lambda_4\pm \lambda)_4,
(\lambda_3 + \lambda_4)_8\}\ ,\nonumber\\
\lambda_1=-\lambda\ ,&& \lambda_2=\lambda_3=\lambda\ ,\qquad
\mu = \ft12\{0_{12}, (2\lambda)_4, (\lambda_4- \lambda)_4,
(\lambda_4 + \lambda)_{12}\}\ ,\label{enhance2f}\\
\lambda_1=-\lambda\ ,&& \lambda_2=\lambda_3=\lambda_4=\lambda\ ,\qquad
\mu = \{0_{16}, \lambda_{16}\}\ ,\nonumber
\eea
where the subscript denotes the degeneracy of each eigenvalue.  Thus these
three cases preserve $\ft14$, $\ft38$ and $\ft12$ of the supersymmetry
respectively, in contrast to $\ft18$ for generic values of the charge
parameters.  Note that in these cases, although the Bogomol'nyi matrices have
no negative eigenvalues when the supersymmetry enhancements occur, the
metrics of the solutions still seem to have naked singularities since one of
the Page charges $\lambda_\a$ is negative.  If we relax the condition that
the eigenvalues of the Bogomol'nyi matrix should be non-negative, then
supersymmetry enhancement can occur for $N=2$ and 3 as well.  For $N=2$, the
solutions can also preserve a fraction $\ft{k}{8}$ of the supersymmetry,
with $k=4$ for appropriately-chosen non-vanishing $\lambda_\a$; for $N=3$,
we can have $k=2$ or 3.  In the case of $N=4$, in addition to the
supersymmetry enhancements described in (\ref{enhance2f}), we can have also
$k=5$ and 6 if negative eigenvalues are allowed, in which case the solutions
preserve more than $\ft12$ of the supersymmetry.\footnote{This does not
violate the classification of supermultiplets given in \cite{fs}, since
non-negativity of the commutator of supercharges was assumed there.  In
fact, if we require that all the eigenvalues of the Bogomol'nyi matrix be
non-negative, then all the solutions preserve no more than $\ft12$ of the
supersymmetry, which is consistent with the classification in \cite{fs}.}

    So far we have discussed elementary solutions, where the field strengths
carry electric charges.  The discussion for the solitonic solutions is
analogous and the conclusions are the same.  In $D=4$ dyonic solutions can
occur, since the dual of a 2-form is again a 2-form.  Although, in our
multi-scalar solutions, all the participating field strengths have the same
purely elementary or purely solitonic nature, some can nevertheless be the
duals of the original ones in the case $D=4$, and therefore, in terms of the
original field strengths, we can have solutions with mixed electric and
magnetic charges.  These were called dyonic solutions of the first type in
\cite{lp}.  In (\ref{eigenvalues2f}), for the cases $N=2$ and 3, we presented
solutions with purely electric charges.  They also exist for purely
magnetic charges, and in $D=4$ they also exist for mixed dyonic charges of
the kind we just discussed.  For the case $N=4$, which occurs only in $D=4$,
the situation is different, in that there are no purely electric or purely
magnetic solutions; they are intrinsically dyonic.

      Before finishing the discussion of the 2-form solutions, we should
remark that there is a further subtlety for the case of $N=4$ scalars. As
was observed in \cite{lp}, the bosonic equations of motion governing the
single-scalar solutions leave the signs of the Page charges for the
participating field strengths undetermined.  If we choose our conventions so
that the mass is always given by $m=\ft14\sum_\a \lambda_\a$, then the
bosonic equations allow solutions where the individual Page charges $P_\a$
can be either $+\ft14\lambda_\a$ or $-\ft14\lambda_\a$.  If we calculate the
eigenvalues of the Bogomol'nyi matrices, we find that in the cases
$N=1,2,3$, they are insensitive to the choices of the signs of the Page
charges.  However, for $N=4$ the signs do matter, and there are precisely
two inequivalent sets of eigenvalues that can arise.  Only one of these
includes zero eigenvalues, and this is the supersymmetric single-scalar
solution with $N=4$ participating field strengths, whose generalisation to 4
scalars we have discussed above.  The other single-scalar solution also
generalises to a 4-scalar solution, which is inequivalent to the one
described above.  For this case, we find that the eigenvalues are
\be
\mu=\ft12 \{\lambda_1,
\lambda_2, \lambda_3, \lambda_4, \lambda_1 + \lambda_2 + \lambda_3,
\lambda_1 + \lambda_2 + \lambda_4, \lambda_1 + \lambda_3 +
\lambda_4,\lambda_2 + \lambda_3 + \lambda_4\}\ ,
\ee
with each eigenvalue having degeneracy 4.  Note that in this case none of
the eigenvalues is proportional to the mass, for generic $\lambda_\a$. This
solution reduces to supersymmetric lower-$N$ cases when any of the
parameters is zero.  Supersymmetry enhancements also occur for certain
non-vanishing Page charges; however, if we require that all the eigenvalues
be non-negative, then this 4-scalar solution is always
non-supersymmetric for all allowed non-vanishing Page charges.

     We now turn our attention to multi-scalar solutions for 1-form field
strengths, which give rise to solitonic $(D-3)$-branes in $D$ dimensions. As
has been shown in \cite{lp}, the number of participating field strengths in
supersymmetric single-scalar solutions can be $N=1,2,\ldots, 7$.  The
solutions with $N=1,2$ and 3 occur in dimensions $D\le 9,8$ and 6
respectively.  There are two inequivalent supersymmetric solutions for
$N=4$, one of which occurs in $D\le 6$ and the other in $D=4$.  The
solutions with $N=5,6$ and 7 all occur in $D=4$ only.   All the solutions
with $N\ge 2$ can be generalised to $N$-scalar solutions, as given in
(\ref{gensol}).   The relevant terms in the Bogomol'nyi matrix for these
solutions are given by
\be
{\cal M} = m\oneone + \ft16 v_{ijk} \Gamma_{\hat1\hat2ijk} + \ft12 q_{ij}
\Gamma_{\hat1\hat2ij} + v^*\, \Gamma_{012} + v_i^*\, \Gamma_{01i}
\ ,\label{bog1f}
\ee
where $v_{ijk}$ and $q_{ij}$ are the Page charges of the field strengths
$F_{\sst{M}ijk}$ and  ${\cal F}_{\sst{M}}^{(ij)}$ coming from the 4-form and
vielbein respectively, $v^*$ is the Page charge of the dual of the
4-form $F_{\sst{MNPQ}}$ (in $D=5$ only), and $v_i^*$ are the Page charges of
the duals of the 3-forms $F_{\sst{MNP}i}$ (in $D=4$ only).  All these Page
charges are magnetic, since the elementary ansatz given in (\ref{fansatz})
does not encompass 1-form field strengths.  We find that the Page charges,
and the eigenvalues of the Bogomol'nyi matrix, for the $N$-scalar solutions
for the 1-forms are given by
\bea
N=2:\!&&\! \{q_{12}, v_{123}\} = \ft14 \{\lambda_1, \lambda_2 \}
\ ,\qquad {\rm for}\,\, D\le 8\ ,\nonumber\\
\mu\!&=&\!\ft12 \{ 0, \lambda_1, \lambda_2, \lambda_{12} \}
\ ,\nonumber\\
N=3:\!&&\! \{q_{12}, q_{45}, v_{123} \} = \ft14 \{ \lambda_1, \lambda_2,
\lambda_3 \}\ ,\qquad {\rm for}\,\, D\le 6\ ,\nonumber\\
\mu \!&=&\! \ft12 \{ 0, \lambda_1, \lambda_2, \lambda_3, \lambda_{12},
\lambda_{13}, \lambda_{23}, \lambda_{123} \}\ ,\nonumber\\
N=4':\!&&\! \{q_{12}, q_{45}, v_{123}, v_{345}\} = \ft14 \{\lambda_1,
\lambda_2,\lambda_3, \lambda_4\}
\ ,\qquad {\rm for}\,\, D\le 6\ ,\nonumber\\
\mu\!&=&\!\ft12 \{ 0, \lambda_{14},\lambda_{24},\lambda_{34}, \lambda_{12},
\lambda_{13}, \lambda_{23}, \lambda_{1234} \}\ ,\nonumber\\
N=4:\!&&\!\{q_{12}, q_{34}, q_{56}, v_{127}\} = \ft14 \{\lambda_1,
\lambda_2,\lambda_3, \lambda_4\}\ ,\qquad {\rm for}\,\, D =4\ ,\nonumber\\
\mu \!&=&\! \ft12\{0, \lambda_1, \lambda_2, \lambda_3, \lambda_4,
\lambda_{12},
\lambda_{13}, \lambda_{14}, \lambda_{23}, \lambda_{24}, \lambda_{34},
\lambda_{123}, \lambda_{124}, \lambda_{134}, \lambda_{234}, \lambda_{1234}\}
\ ,\label{eigenvalues1f}\\
N=5:\!&&\! \{q_{12}, q_{34}, q_{56}, v_{127}, v_{347} \} =
\ft14 \{\lambda_1, \lambda_2, \lambda_3, \lambda_4 \}\ ,
\qquad {\rm for}\,\, D =4\ ,\nonumber\\
\mu\!&=&\! \ft12\{0, \lambda_3, \lambda_{12}, \lambda_{14}, \lambda_{15},
\lambda_{24},\lambda_{25},\lambda_{45},\lambda_{345},\lambda_{235},
\lambda_{234},\lambda_{135},\lambda_{134}, \lambda_{123},
\lambda_{1245},\lambda_{12345} \} \ ,\nonumber\\
N=6:\!&&\! \{q_{12}, q_{34}, q_{56}, v_{127}, v_{347}, v_{567} \}=
\ft14 \{\lambda_1, \lambda_2, \lambda_3, \lambda_4, \lambda_5, \lambda_6 \}
\ ,\qquad {\rm for}\,\, D =4\ ,\nonumber\\
\mu \!&=&\! \ft12\{0, \lambda_{14}, \lambda_{25}, \lambda_{36},
\lambda_{123}, \lambda_{126}, \lambda_{135}, \lambda_{156},
\lambda_{234},  \lambda_{246}, \lambda_{345},
\lambda_{456}, \lambda_{1245},
\lambda_{1346}, \lambda_{2356},  \lambda_{123456} \}\nonumber\\
N=7:\!&&\! \{q_{12}, q_{34}, q_{56}, v_{127}, v_{347}, v_{567}, v^*_7\}
=\ft14 \{\lambda_1, \lambda_2, \lambda_3, \lambda_4, \lambda_4, \lambda_5,
\lambda_6, \lambda_7 \}\ ,\qquad {\rm for}\,\, D =4\ ,\nonumber\\
\mu\!&=&\!\nonumber\\
\{0,&&\!\!\!\!\!\!\!\!\!\!\!\!\!\!
\lambda_{126}, \lambda_{135}, \lambda_{234},
\lambda_{147},
\lambda_{257}, \lambda_{367}, \lambda_{456}, \lambda_{1245},
\lambda_{1346},
\lambda_{2356}, \lambda_{1567}, \lambda_{2467}, \lambda_{3457},
\lambda_{1237},
\lambda_{1234567} \}\ .\nonumber
\eea
Here, for convenience, we have defined $\lambda_{\a\beta\cdots\gamma}=
\lambda_\a + \lambda_\beta + \cdots +\lambda_\gamma$.   Note that we
preseneted only one representative set of Page charges among many
possibilities for each case, since they have the identical eigenvalues. In
each case, the degeneracy of each eigenvalue for a particular solution is
the same, with the total number of eigenvalues being 32.  The last
eigenvalue in each case is twice the mass of the corresponding solution. The
eigenvalues of a higher-$N$ case reduce to those of all the lower-$N$ cases
when certain of the $\lambda_\a$'s are set to zero, and hence we shall only
consider solutions with non-vanishing $\lambda_\a$.  When $N=5$, there are
two inequivalent ways of reducing to $N=4$, giving the two cases that we
denote by $N=4$ and $N=4'$.   Note that the eigenvalues for the first three
cases are identical to those for the 2-form field strengths that we have
discussed previously.

      For generic values of $\lambda_\a$, the solutions preserve $2^{-N}$ of
the supersymmetry for $N=1,2,3,4$.  The $N=4'$ solutions preserve $\ft18$ of
the supersymmetry and all the $N=5,6,7$ solutions preserve $\ft1{16}$.  It
is easy to verify that when all $\lambda_\a$ are equal, the solutions reduce
to the single-scalar solutions with $\Delta = \ft{4}{N}$, which were
discussed in \cite{lp}.  The single-scalar solutions preserve the same
fractions of supersymmetry as their generic multi-scalar extensions.  For
certain choices of $\lambda_\a$, the multi-scalar solutions can have
supersymmetry enhancement.  In particular, it occurs when the mass $m=\ft14
\sum_\a \lambda_\a$ is zero; however, as in the case of 2-form field
strengths, the Bogomol'nyi matrices for the solutions then have indefinite
signature.  If we require that all the eigenvalues be non-negative, then for
the cases $N=2,3$ and 4 there is no supersymmetry enhancement when the
$\lambda_\a$'s are non-vanishing. The analysis of the supersymmetry
enhancement for the case $N=4'$ is equivalent to the $N=4$ case for 2-form
field strengths, which we have already discussed.  For $N=5,6$ and 7, there
are many ways to choose the parameters to achieve supersymmetry enhancement,
and we shall not present all of them. Choosing the parameters $\lambda_\a$
appropriately, the $N=5$ solutions can preserve $\ft{k}{16}$ of the
supersymmetry with $k=2,3,4$; similarly $k=2,3,4,5$ or 6 for the $N=6$ case;
and $k=2,3,4,5,6$ or 7 when $N=7$, in contrast to $k=1$ for generic values
of $\lambda_\a$ for each of these values of $N$.  If we relax the condition
that all the eigenvalues of the Bogomol'nyi matrices be non-negative,
further supersymmetry enhancements can occur, which includes the massless
$p$-brane solutions.  From (\ref{eigenvalues1f}), it is straightforward to
obtain the possible fractions of the supersymmetry that a solution can
preserve, and we shall not present the details; the maximal fractions for
non-vanishing Page charges turn out to be $\ft12$, $\ft38$, $\ft34$,
$\ft38$, $\ft7{16}$, $\ft58$ and $\ft12$ for $N=2,3,4',4,5,6$ and 7
respectively.

     As in the case of 2-form solutions, the eigenvalues of the Bogomol'nyi
matrix are sensitive to the signs of the Page charges for certain 1-form
solutions.  If we choose our conventions so that the mass is always given by
$m =\ft14\sum_\a \lambda_\a$, then bosonic equations allow solutions where
the individual Page charges $P_\a$ can either be $+\ft14 \lambda_\a$ or
$-\ft14 \lambda_\a$.  For the 1-form solutions with $N=1,2,3$ and 4, the
eigenvalues are insensitive to the choice of the signs of the Page charges.
However, for the $N=4',5,6$ and 7 cases, the signs do matter, and there are
precisely two inequivalent sets of eigenvalues that can arise.  One of
these, which includes zero eigenvalues for generic $\lambda_\a$, is
presented in (\ref{eigenvalues1f}). The other does not have zero eigenvalues
for generic Page charges.  For $N=7$, we find that the eigenvalues are given
by
\bea
\mu &=& \ft12\{\lambda_{7}, \lambda_{14}, \lambda_{25}, \lambda_{36},
\lambda_{123}, \lambda_{156}, \lambda_{246}, \lambda_{345},
\lambda_{1267}, \lambda_{1357}, \lambda_{2347}, \lambda_{4567},\nonumber\\
&&\,\quad\lambda_{12457}, \lambda_{13467}, \lambda_{23567},
\lambda_{123456}\}\ .
\eea
It reduces to the $N=6,5$ and $4'$ cases when $\lambda_4$, $\lambda_5$ and
$\lambda_6$ are successively set to zero.  In each of these four cases, none
of the eigenvalues is proportional to the mass, for generic $\lambda_\a$.
When all $\lambda_\a$'s are set equal, the solutions reduce to single-scalar
solutions that break all the supersymmetry.  If we require that all the
eigenvalues be non-negative, the $N=4'$ solutions will always be
non-supersymmetric for all non-vanishing $\lambda_\a$.  However, for the
$N=5,6$ and $7$ cases, there can be supersymmetry enhancement for certain
non-vanishing choices of $\lambda_\a$.  Choosing the parameters $\lambda_\a$
appropriately, the $N=5$ solutions can preserve $\ft{k}{16}$ of the
supersymmetry with $k=1,2,3,4$; similarly $k=1,2,3,4,5,6$ for $N=6$ and
$k=1,2,3,4,5$ for $N=7$.  Thus although the corresponding single-scalar
solutions are non-supersymmetric, their multi-scalar generalisations can be
supersymmetric.  Further supersymmetry enhancements can occur if we relax
the condition that all the eigenvalues of the Bogomol'nyi matrix be
non-negative.  It is straightforward to enumerate the possibilities from the
eigenvalues given above.

     To conclude, in this paper we have shown that the supersymmetric
single-scalar $p$-brane solutions in the maximal supergravity theories with
$N$ participating field strengths for $N=1,2, \ldots, 7$ can be extended to
$N$-scalar solutions. The $N$ Page charges, which were equal in the
single-scalar solutions, become independent free parameters.  For the 4-form
and 3-form field strengths, $N=1$; for the 2-form field
strengths $N=1,2,3,4$; and for the 1-form field strengths $N=1,2, \ldots,
7$.  We summarise the 2-form and 1-form solutions in the following table:
\bigskip

\centerline{
\begin{tabular}{|c||c|l||c|c|}\hline
{\phantom{D}Dim.\phantom{D}}&\multicolumn{2}{c||}
{\phantom{DDDDD} 2-Forms \phantom{DDDDD}} &
\multicolumn{2}{c|}{\phantom{DDDDD}1-Forms\phantom{DDDDD}} \\ \hline\hline
$D=10$& $\phantom{D}N=1\phantom{D}$ &\phantom{DD}$p=0,6$ & &  \\ \hline
$D=9$ & $N=2$ &\phantom{DD}$p=0,5$ & $N=1$ & $p=6$ \\ \hline
$D=8$ &       &\phantom{DD}$p=0,4$ &$N=2$&$p=5$ \\ \hline
$D=7$ &       &\phantom{DD}$p=0,3$ &     &$p=4$ \\ \hline
$D=6$ &       &\phantom{DD}$p=0,2$ &$N=3,4'$&$p=3$ \\ \hline
$D=5$ & $N=3$ &\phantom{DD}$p=0,1$ &        &$p=2$ \\ \hline
$D=4$ & $N=4$ &\phantom{DD}$p=0$   &$N=4,5,6,7$&$p=1$ \\ \hline
\end{tabular}}
\bigskip

\centerline{Table 1: Multi-scalar $p$-brane solutions}
\bigskip

\noindent  Here we list the highest dimensions where $p$-brane solutions
with the indicated numbers $N$ of field strengths first occur.  They then
occur also at all lower dimensions.  All these solutions preserve certain
fractions of the $D=11$ supersymmetry for generic values of the Page
charges.  Further supersymmetry enhancement can occur for certain choices of
non-vanishing Page charges, which sometimes gives rise to solutions whose
Bogomol'nyi matrix has indefinite signature. In some cases, however,
supersymmetry enhancement can occur while still avoiding negative
eigenvalues in the Bogomol'nyi matrix.  For all the multi-scalar $p$-brane
solutions, the mass is equal to the sum of the Page charges. Thus in
principle it is possible to have massless $p$-brane solutions.  In fact,
when the mass is set to zero the supersymmetry of the solutions is enhanced.
However, the Bogomol'nyi matrix will have an indefinite signature. Since
supersymmetry enhancement occurs only when some of the Page charges are
negative, it follows from (\ref{gensol}) that the metric seems to have naked
singularity regardless of whether the Bogomol'nyi matrix has non-negative
eigenvalues or not.  It seems that all the Page charges have to be positive
in order to ensure that the metric does not have any naked singularities.
The status of the naked singularities in these solutions is unclear. On the
other hand, we would certainly expect that, owing to the quantum positivity
argument, only solutions with non-negative eigenvalues of the Bogomol'nyi
matrix are acceptable.  In particular, this seems to cast doubt on the
validity of the massless super $p$-brane solutions in the spectrum of the
theory.

    In this paper, we have been primarily concerned with purely elementary
or purely solitonic multi-scalar solutions.  In $D=6$ and $D=4$, dyonic
solutions can also occur.   There are two different types of dyonic
solutions. In dyonic solutions of the first type, each individual field
strength carries either electric or magnetic charge but not both. On the
other hand, in dyonic solutions of the second type, each individual field
strength carries both electric and magnetic charges.  In $D=6$, one can only
construct dyonic solutions \cite{dfkr} of the second type, with just one
3-form field strength involved \cite{lp}, and hence there is no multi-scalar
extension in this case.  In $D=4$, one can construct dyonic solutions of
both the first and second types.  We presented the multi-scalar
generalisations of single-scalar dyonic solutions of the first type for all
$N=2,3$ and 4 field strength cases.  There are two single-scalar dyonic
solutions of the second type in $D=4$, with $N=2$ and 4 \cite{lp}.  It would
be of interest to generalise these to multi-scalar solutions.

     For the non-supersymmetric single-scalar solutions discussed in
\cite{lp}, the dot products of the dilatonic vectors $\vec a_\a$ do not
satisfy (\ref{mmatrix}), and thus multi-scalar solutions of the kind we have
discussed in this paper cannot occur.   In fact, the multi-scalar solutions
for these cases have a much more complicated form. Although the metrics of
all these solutions can still be asymptotically Minkowskian as
$r\longrightarrow \infty$, the metric structure will become very complicated
near the origin. We showed that for the supersymmetric multi-scalar
solutions, owing to the fact that the Page charges become independent,
supersymmetry enhancement can occur for appropriately-chosen Page charges.
It would be of interest to know whether the multi-scalar generalisations of
the non-supersymmetric single-scalar solutions can also become
supersymmetric for certain choices of Page charges.

\section*{Acknowledgement}

     We are grateful to M.J.~Duff for useful discussions.

\pagebreak

\end{document}